\documentclass[seceq,preprint]{ptptex}
\usepackage{wrapft}
\usepackage{graphicx}

\newcommand{\eq}{Eq.\ }
\newcommand{\eqs}{Eqs.\ }

\newcommand{\id}{{\cal I}}
\def\halftext{.471\textwidth}

\preprintnumber[3cm]{
hep-th/0405097}

\title{
Worldsheet Geometry of 
Classical Solutions in String Field Theory
}

\author{
Syoji \textsc{Zeze}
\footnote{E-mail: zeze@asuka.phys.nara-wu.ac.jp}
}

\inst{
Department of Physics, Nara Women's University, Japan
}

\abst{
We investigate classical solutions of string 
field theory proposed by Takahashi and Tanimoto 
in the case of even order polynomial functions.
The modified BRS charge and the Feynman
propagator of open string field theory expanded about
the solution are specified by the
Jenkins-Strebel quadratic differential,
which describes the geometry of the string worldsheet.
We show that the solution becomes nontrivial
when two second order poles of the quadratic differential coincide
on the unit disk.
}

\begin{document}

\maketitle

\section{Introduction}

Classical solutions in string field theory (SFT) have been
extensively investigated since 
Sen and Zwiebach\cite{rf:SZ} demonstrated that 
the Lorentz invariant classical solution of 
cubic open string field theory\cite{rf:CSFT}
corresponds to 
the nontrivial vacuum of the tachyonic
potential.
They evaluated the solution and 
the D25-brane tension using the level 
truncation scheme,~\cite{rf:KS} 
which truncates the string field at finite mass level.
Surprisingly, their estimation of the tension
with the level (4,8) approximation is within 1\% of the expected
value. 
Following their successful result, many works  
using the level truncation scheme have been carried out~\footnote{%
For example, see references in Ref.~\citen{rf:review}. } 
and they support Sen's conjecture~\cite{rf:sencond}.

However, the level truncation analysis is insufficient
to prove and understand Sen's conjecture completely.
In particular, 
the worldsheet geometry behind the physics of tachyon condensation 
is  lost when the string field is truncated  at a finite level,
because  an infinite number of degrees of freedom
are needed to describe  string propagation on a worldsheet,
which is a  Riemann surface.  
One possible way to treat such problems is 
to consider an exact solution of equation of the motion of SFT, 
because it contains infinitely many degrees of freedom 
that are sufficient to
describe a string worldsheet.

One attempt to construct a framework in which to treat 
the above-mentioned problem is represented
by vacuum string field theory 
 (VSFT).\cite{rf:vsft,rf:vsftsol,rf:vsfthalf,rf:vsftbcft,rf:vsftghost} 
The authors of Ref.~\citen{rf:vsft} proposed a pure ghost
BRS charge as a kinetic operator of SFT after tachyon condensation,
instead of finding an exact classical solution. 
This BRS charge ensures the disappearance of string excitations 
because it has a trivial cohomology.
The 
kinetic operator in the Siegel gauge, obtained by anticommuting
the BRS charge with $b_0$, is merely 
a {\it c}-number that has no geometrical meaning. Therefore 
the authors of Ref.~\citen{rf:vsftghost}
 proposed a regularization procedure using the kinetic operator
$L_0$, which represents open string propagation on a flat
worldsheet. They also made the observation that a closed string amplitude 
can be obtained by calculating a correlation function of
the gauge invariant closed string operator\cite{rf:Zwiebach-closedvertex} 
with the regularized propagator. Because the geometry of 
the worldsheet is  same as in the case of ordinary open string 
theory, the effect of tachyon condensation is due to the 
regularization procedure. 
However, it is still not clear whether such a prescription is correct, 
as there is no principle to determine 
appropriate regularization procedure.

Another approach based on exact classical solutions 
is proposed in Ref.~\citen{rf:TT}.
The classical solutions are called `universal solutions',
because they belong to the universal subspace of
the open string Hilbert space\cite{rf:SenUniv}. 
In contrast to VSFT, the modified BRS charge obtained by
expanding the SFT action around about the universal solution
depends on the matter part of open string conformal
field theory. 
Therefore the disappearance of open strings is a quite 
nontrivial problem in this approach.
The solution is not unique but 
can be written in terms of a wide variety of functions
on the complex plane. 
Each solution is uniquely specified by this function.
In Ref.~\citen{rf:TT} 
the simplest solution with one real parameter
is considered. It was found that at a boundary of the parameter space
this solution is nontrivial, and at an interior of 
the parameter space it  becomes a pure gauge solution. 
Subsequently, the cohomology of
the modified BRS charge was investigated in 
Refs.~\cite{rf:KT} and \cite{rf:TZ}. There it was shown that 
at a boundary of parameter space the cohomology
is trivial, and hence there is no open string.

After these works, Drukker made the important observation that
the propagator obtained from a nontrivial universal solution
generates a worldsheet whose boundary shrinks to a 
point\cite{rf:Drkr1,rf:Drkr2}. 
This mechanism can account to the disappearance of the open string
from  the worldsheet geometry. Actually, in Ref.~\citen{rf:Drkr1}
the worldsheet geometry of a purely closed string amplitude 
is studied.

In this paper, we elaborate on such geometrical 
treatment of string propagation by considering
a class of universal solutions composed of polynomial 
functions. We identify the function 
specifying each solution as 
a quadratic differential on the complex plane. 
Quadratic differentials were used extensively in the development of 
the perturbation 
theory of SFT\cite{rf:gidd,rf:GM,rf:GMW,rf:zwclosed,rf:naito}. 
Feynman diagrams studied then are 
obtained by connecting a flat strip or cylinder
whose geometry is locally trivial.
A quadratic differential was used to understand the moduli space 
of Feynman diagrams with many propagators and vertices.
Now our interest is in a single propagator deformed 
by a classical solution of string field theory.
It has a locally nontrivial geometry
represented by poles or zeros of the quadratic differential.

The paper is organized as follows. In \S 2 we review basic facts
concerning universal solutions\cite{rf:TT}. 
In particular, we focus on an algebraic property
of the space of functions labeling each solution.
In \S 3 we define the subspace of universal solutions
composed of even-order polynomial functions explicitly, 
and we show that 
the solution is nontrivial when the zeros of the corresponding function
are on the unit circle.  
In \S 4, we show that 
the function labeling each universal solution defines
a quadratic differential describing the world sheet geometry, 
and plot some examples of trajectory diagrams.

\section{Universal solutions}
\label{sec:universal}

A universal solution\cite{rf:TT} is a 
background independent, Lorentz
invariant classical solution of
cubic open string field theory\cite{rf:CSFT}. 
Such a solution is given by
\begin{equation}
 \Psi_0 = Q_L (F) \id - C_L \left(\frac{(\partial F)^2 }{1+F^2}  \right) \id,
\label{eq:univsol}
\end{equation}
where $\id$ is the identity string field. $F$ is a function 
defined on the complex plane. It satisfies the conditions
\begin{equation}
 F(-1/w) = F(w), \quad F(\pm i) = 0.\label{eq:Fcond}
\end{equation}
The quantities $Q_L (F)$  and $C_L (G)$ are
defined by\footnote{We omit the factor of $1/2\pi i$  in 
 contour integrals.}
\begin{equation}
 Q_L (F) = \int_{\gamma_L} dw \, F(w) J_B (w),\quad
 C_L (G) = \int_{\gamma_L} dw \, G(w) c (w), 
\end{equation} 
where the path $\gamma_L$ 
traverses in the counterclockwise direction on the
left half of the unit circle. $J_B (w)$ and $c(w)$
are the BRS current and conformal ghost, respectively.
We analyze the modified BRS charge~\footnote{In general, 
the modified BRS charge is not
a BRS charge of some conformal invariant
CFT obtained from the BRS gauge fixing. In the rest of this paper,
we shall omit the term `modified' for simplicity.    }
obtained by shifting the string field around the solutions 
instead of analyzing  \eq (\ref{eq:univsol}) directly, 
as this is technically simpler.
Rewriting $F(w)$ as $g(w) -1$,
we can compute the BRS charge as\cite{rf:TT}
\begin{eqnarray}
 Q_g \Psi & = &  Q_B \Psi + \Psi_0 * \Psi + \Psi * \Psi_0 \nonumber   \\  
 &  = &\left[ Q\left(g\right) - C\left(\frac{(\partial g)^2}{g}   \right)
 \right] \Psi,  \label{eq:BRSdef}
\end{eqnarray}
where $Q(f)$ and $C(f)$ are defined by integrals over the unit circle
$\gamma$ as
\begin{equation}
 Q(f)        =  \oint_{\gamma} dw \, f(w) J_B (w), 
\quad C(f)   =  \oint_{\gamma} dw \, f(w)\, c (w). \label{eq:Q(F)def} 
\end{equation}
From the condition (\ref{eq:Fcond}), $g(w)$ must satisfy
\begin{eqnarray}
  g \left(-\frac{1}{w} \right) & = & g(w),\label{eq:gcond1} \\
  g \left(\pm i \right) & = & 1. \label{eq:gcond2}
\end{eqnarray}

We refer to a function $g(w)$ satisfying 
\eqs (\ref{eq:gcond1}) and (\ref{eq:gcond2})
as a {\it universal function}, because it specifies 
the universal solution and the BRS charge uniquely.
It is used frequently in the following. 
Equation (\ref{eq:BRSdef}) allows us to obtain various
BRS charges by choosing the universal function. 
Note that the universal functions 
form an Abelian group 
with respect to ordinary 
multiplication of functions. More
precisely, let $\mathcal{F}$ be the set of all universal functions.
Then it is easy to see that
\begin{equation}
 f, g \in\mathcal{F}  \rightarrow fg \in \mathcal{F}, 
\end{equation}
using \eqs (\ref{eq:gcond1}) and (\ref{eq:gcond2}).
Furthermore, the identity element and the inverse  
are given by $g(w)=1$ and $1/g(w)$, respectively. Thus
$\mathcal{F}$ is an Abelian group with respect 
to ordinary multiplication.

As shown in Refs.~\cite{rf:TT} and \cite{rf:KT}, there exists
a homomorphism from $\mathcal{F}$ to the space of
field redefinitions acting on the string field. To see this,
let us define the conserved ghost current $q(h)$ as
\begin{equation}
 q(h) = \oint_{\gamma} dw \, h(w) \left(J_{gh} (w)- \frac{1}{w}\right),
\end{equation}
where $J_{gh} (w) = :c(w)b(w):$ is 
the ghost number current.\footnote{$q(h)$ is formulated 
to be conserved  on the $N$-string
vertex of SFT by subtracting the ghost number anomaly. }
Using this current, we
can construct a field redefinition operator as
\begin{equation}
 E_g =e^{q(\log g)}.\label{eq:redef}
\end{equation}
Applying 
this operator to the BRS charge (\ref{eq:BRSdef}),
we have the identity given in Ref.~\citen{rf:KT},
\begin{equation}
 E_f Q_g E_{f}^{-1} = Q_{fg}. \label{eq:Q-redef}
\end{equation}
From the above equation, we easily show that
\begin{equation}
 E_f E_g = E_{fg}.
\end{equation}
Thus, the homomorphism mentioned above 
is given by $ g \rightarrow E_g$. 
Moreover, this Abelian group 
is also homomorphic to the Abelian subgroup of the gauge group of SFT. 
Now, recall that  the universal solution (\ref{eq:univsol}) can be 
rewritten as the pure gauge expression
\begin{equation}
 \Psi_{0} = U_g * Q_B {U_g}^{-1},
\end{equation}
where $U_g = e^{q_L (\log g) \id}$
is an element of 
the gauge group of SFT,\footnote{An exponential in $U_g$
is constructed from star products:
$U_g = \id + q_L (\log g) \id +1/2
\left\{q_L (\log g) \id  \right\} * \left\{q_L (\log g) \id  \right\} 
+ \cdots$.
} 
and $q_L (\log g)$ is 
an integral of $\log g (w) J_{gh}(w)$ on the left half of 
the unit circle.\cite{rf:TT}  Using formulas given in Ref.~\citen{rf:TT},
we can see that 
\begin{equation}
 U_f * U_g = U_{fg}
\end{equation}
holds.  The above equation gives a homomorphism from the space
of universal functions $\mathcal{F}$ to the subspace of the SFT 
gauge group generated by $U_g$.
Therefore, we can determine the structure of this subgroup
by analyzing $\mathcal{F}$. 

In order to classify elements of $\mathcal{F}$, it is 
convenient to focus on the operator $E_g$. 
It acts on the string field as a field redefinition 
given by $\Psi' = E_g \Psi$. 
Using the fact that $E_g$ leaves string
field vertices invariant\cite{rf:TT}, we can show 
the formal identity 
\begin{equation}
 S[Q_B, \Psi] = S[Q_g, \Psi'],\label{eq:actionequiv}
\end{equation} 
where
\begin{equation}
S[Q_g, \Psi] = \int \frac{1}{2} \Psi * Q_g \Psi +
\int \frac{1}{3} \Psi * \Psi *\Psi
\end{equation}
is the cubic open SFT action\cite{rf:CSFT} with 
BRS operator $Q_g$. 
If $E_g$ is a regular transformation on the 
string field, \eq (\ref{eq:actionequiv}) 
implies that SFT with the BRS operator $Q_g$ is equivalent to the
original theory, and the classical solution labeled by $g(w)$
is merely a pure gauge solution.  However, it happens that
$E_g$ becomes singular and the transformation on the string
field is ill-defined, because
it is an  infinite-dimensional linear transformation on the
component fields of $\Psi$. For example, such a singularity can be
seen from the normal ordering constant of $E_g$:
\begin{equation}
 E_g = N :E_g:.  
\end{equation}
If $N$ diverges, the transformation $E_g$ is singular and ill-defined,
and the corresponding classical solution is expected 
to be nontrivial. 
Therefore, we define the singular transformation
$E_g$ as a transformation whose normal ordering constant is divergent,
and the nontrivial element of $\mathcal{F}$ as
the universal function $g(w)$ associated with a singular transformation.  

Our goal is to obtain 
all singular solutions up to regular field redefinitions and to classify
them. This is accomplished as follows. First, let $\mathcal{F}_r$ be
a subgroup of $\mathcal{F}$ giving a regular transformation.
Then, the space of inequivalent solutions is the coset
\begin{equation}
 \mathcal{K} = \mathcal{F}/ \mathcal{F}_r.\label{eq:coset}
\end{equation}
In particular, the identity element of $\mathcal{K}$ 
corresponds to the ordinary BRS charges $Q_B$. Other elements
represent BRS charges that are equivalent neither to $Q_B$ nor
to each other.

It is surprising that a classification of exact solutions of SFT
--- which involves the complicated structure of the star product
in general ---  reduces to the significantly simpler problem of
determining the 
Abelian group of the multiplication of universal functions.   

\section{Even finite universal solutions}

Let us consider a universal function containing 
finite powers of $w$.  It can be decomposed as
\begin{equation}
  g(w)   =  g_{+} (w) + g_{-} (w),\label{eq:oddeven}
\end{equation}
where $g_{+} (w)$ and $g_{-} (w)$ are
even and odd functions of $w$, respectively.
Imposing  \eq (\ref{eq:gcond2}) on 
\eq (\ref{eq:oddeven}), 
we see that these functions must satisfy the relation
\begin{equation}
 g_{+} (\pm i) =1, \quad g_{-} (\pm i) =0.
\end{equation}
We can consider the subset of universal functions satisfying
 $g_{-} (w)=0$, as in this case $g(w)$ still satisfies
the condition (\ref{eq:gcond2}) and actually belongs to
the subgroup of $\mathcal{F}$ generated by even polynomial functions.
In the following, we limit ourself to this case, because 
$g(w)$ has a simpler structure than 
that in the case involving the odd part\footnote{
When the odd part is involved, some Laurent coefficients of $g(w)$ 
become purely imaginary.}. 
In this case, the Laurent expansion of $g(w)$ is given by 
\begin{equation}
 g(w) = \sum_{n=0}^{N} a_n \left(w^{2n} + w^{-2n} \right), \label{eq:g-laurant}
\end{equation}
where $N$ is a positive integer and 
we have used \eq (\ref{eq:gcond1}). 
It is convenient to rewrite  
\eq (\ref{eq:g-laurant}) into the rational form 
\begin{equation}
g (w) = \frac{P_{2N} (w^2)}{w^{2N}}, 
\end{equation}
where $P_{2N}$ is a $2N$th order polynomial
of $w^2$. From \eq (\ref{eq:gcond1}), it is clear  
 that if $X$ is a zero of $P_{2N}$, then
$X^{-1}$ also must be a zero of $P_{2N}$. 
Thus, the universal function can be expressed as 
\begin{equation}
 g(w) = \lambda \prod_{k=1}^{N} \frac{
\left(w^2 -X_k \right)\left(w^2 - {X_k}^{-1}\right)     }{w^{2}}, 
\label{eq:gprod}
\end{equation}
where $X_k \ (k=1,\cdots , N)$ are  complex parameters. Because 
$X_k$ and ${X_k}^{-1}$ appear in pairs, we can
set $|X_k| \leq 1$ 
without loss of generality.
The constant $\lambda$ in \eq (\ref{eq:gprod}) is
determined from \eq (\ref{eq:gcond2}) as
\begin{equation}
 \lambda =  \prod_{k=1}^{N} 
\frac{-1}{\left(1+X_k \right) \left(1+{X_k}^{-1}\right)}.\label{eq:lambdadef} 
\end{equation} 
Note that \eq (\ref{eq:gprod}) can be
represented as a product of `$N=1$'
universal functions as
\begin{equation}
 g(w)  = \prod_{k=1}^{N} g_{X_k} (w),\label{eq:decomposeg} 
\end{equation}
where
\begin{eqnarray}
g_{X_k} (w) & = & \lambda_k 
\frac{\left(w^2 -X_k \right)\left(w^2 - {X_k}^{-1}\right)     }{w^{2}},\label{eq:gXk} \\
\lambda_k & = &\frac{-1}{\left(1+X_k \right) \left(1+{X_k}^{-1}\right)}.\label{eq:lmdk}
\end{eqnarray}

A further condition 
must be imposed on 
the set of parameters $\{X_k\}$ that results from the
Hermicity of the BRS charge. 
As shown in Appendix \ref{sec:hermite},
the Hermicity of $Q_g$ is equivalent to $g(w)$ being purely real
on the unit disk. 
From \eqs (\ref{eq:decomposeg}) -- (\ref{eq:lmdk}),
we find that
\begin{equation}
 \overline{g(w)}  = \prod_{k=1}^{N} g_{\overline{X_k}} (w) \label{eq:complexg}
\end{equation}
is satisfied on the unit disk. Then, from \eqs 
 (\ref{eq:decomposeg}) and (\ref{eq:complexg}),
the Hermicity condition reads
\begin{equation}
 \overline{X_k} = X_{\sigma (k)} \quad (k=1, 2,  \cdots , N).\label{eq:perm}
\end{equation}
Here, $\sigma$ is a permutation that sends $\{1, 2, \cdots , N \}$
to $\{\sigma (1), \sigma (2), \cdots , \sigma (N)  \}$. 
 Thus we have obtained the class of universal solutions defined by
\eqs (\ref{eq:gprod}), (\ref{eq:lambdadef}) and (\ref{eq:perm}).
We call the solutions in this class  `even finite universal solutions'.
Each solution is labeled by a set of parameters 
$\{ X_k \}$ which satisfies  \eq  (\ref{eq:perm}).

Now that the even finite solutions are defined explicitly, 
we can specify nontrivial solutions among them 
by carrying out the normal ordering defined by \eq (\ref{eq:redef})
on the field redefinition operator $E_g$. If we write $h(w) = \log g(w)$, 
this operation is expressed as
\begin{equation}
 e^{q (h)} = e^{\frac{1}{2}[{q^{+}(h)},
{q^{-}(h)}    ]}:e^{q (h)}:,
\end{equation}
where ${q^{+}(h)}$ and ${q^{-}(h)}$ are the positive and negative
frequency parts of $q(h)$, respectively. Using 
formulas given in Appendix \ref{seq:qh}, we obtain 
\begin{equation}
 E_g = \left[
\prod_{k=1}^{N}
\frac{1}{1-X_{k}^{2}}
\right]
\left[
\prod_{k<l}^{N}
\frac{1}{(1-X_k X_l)^2}
\right]
:E_g: .\label{eq:evennormal} 
\end{equation}
Because  $|X_k|\leq 1$, the factor
$1-X_k X_l$ in \eq (\ref{eq:evennormal})
becomes zero if and only if both  $X_k$ and $X_l$ 
are on the unit circle. Furthermore, the quantity $1-{X_k}^2$ in \eq
(\ref{eq:evennormal}) becomes zero when $X_k =\pm 1$, 
where $X_k$ is also 
on the unit circle. 
Therefore,
$E_g$ becomes a singular field redefinition
if at least one of the elements of $\{X_k\}$ lies on the
unit circle. Because a  regular $E_g$ corresponds to
trivial pure gauge transformation\cite{rf:TT}, 
we conclude that {\it an
 even finite  universal solution is nontrivial if and only if
some zeros of $g(w)$\footnote{Note that $(X_k)^{1/2}$ is a zero 
of $g(w)$. } lie on the unit circle.
}
 
Our next task is 
to obtain an `irreducible' nontrivial solution
by removing the regular part from singular solution. For example, 
consider 
a nontrivial even finite solution given by 
a universal function $g(w)$ that has
some zeros on the unit circle. 
Suppose that there exists the factorization
\begin{equation}
 g(w)= g_r (w) \tilde{g} (w), \label{eq:fact}
\end{equation}
where $g_r (w)$ and $\tilde{g} (w)$ are even finite 
universal functions such that  
$g_r (w)$ has all zeros inside the unit disk whereas $\tilde{g} (w)$
has some zeros on the unit circle.  Then, using \eq (\ref{eq:Q-redef}), 
we can remove $g_r (w)$ through a regular field redefinition and 
thereby obtain the 
reduced universal function $\tilde{g} (w)$. 

Though we have assumed the factorization (\ref{eq:fact})
, at this stage, we do not  know whether it is always possible.
In the following, we prove that in fact it is the case.  
To prove the factorization, 
we must take into
account the Hermicity condition (\ref{eq:perm}).
First, it is useful to consider  
a special element of $\mathcal{F}$ such that
the permutation of \eq (\ref{eq:perm}) is {\it cyclic}. 
In this case, we can set
\begin{eqnarray}
 \overline{X_k} & = & X_{k+1}, \quad (k=1, \cdots, N-1 ),\label{eq:cyc1} \\
 \overline{X_N} & = & X_{1},\label{eq:cyc2} 
\end{eqnarray} 
without loss of generality. From \eqs (\ref{eq:cyc1}) and
(\ref{eq:cyc2}), it is clear that the set $\{X_k\} \ (k=1,\cdots ,N)$ 
can be expressed as
\begin{equation}
\{ 
\underbrace{X_1, \overline{X_1}, X_1, \overline{X_1},  \cdots  }_{N}
\}, \label{eq:manyX}
\end{equation}
and the universal function  has the expression
\begin{equation}
 g^{c}_{X_1}(w) = \left\{g_{X_1} (w)\right\}^{n}
\left\{g_{\overline{X_1}} (w)\right\}^{N-n},\label{eq:cycfun}
\end{equation}
where the superscript $c$ denotes ``cyclic'' and  $n$ is given by 
\begin{equation}
 n = \begin{cases}
      \frac{N}{2} & N \in 2 \mathbb{N}, \\
      \frac{N-1}{2} & N \in 2 \mathbb{N} -1. \\
     \end{cases}
\end{equation}
From \eq (\ref{eq:manyX}), we find that $g^{c}_{X_1}$ must be
either irreducible or entirely regular, because it 
depends on only the single parameter $X_1$. 

Let us now return to the case of general
even finite universal functions with 
the permutation $\sigma$. 
It is well known that any 
permutation can be written as a 
direct sum of cyclic permutations.
This implies that
$g(w)$ factorizes into a product of `cyclic' functions defined by
\eq (\ref{eq:cycfun}) as 
\begin{equation}
 g (w) = g^{c}_{X_1} (w) g^{c}_{X_2} (w) \dotsb g^{c}_{X_M} (w),\label{eq:cyclicdecomp}
\end{equation} 
where $M$ is the number of cyclic permutations contained in $\sigma$.
If  $X_k$ is inside the unit disk,
we can remove $g^{c}_{X_k} (w)$ by applying the field redefinition operation
on the BRS operator.
 Let us assume that the elements of $\{X_1, X_2, \cdots , X_m\} \ (m < M)$ are 
located on the unit circle and other zeros are inside unit disk.
Then, $g(w)$ factorizes into singular and nonsingular parts 
as in \eq (\ref{eq:fact}). Removing the regular part from $g(w)$, 
we obtain the irreducible universal function
\begin{equation}
 \tilde{g}(w) = g^{c}_{X_{1}} (w) \dotsb g^{c}_{X_{m}} (w),\label{eq:nontrivfinal}.
\end{equation}
Therefore, we conclude that
{\it 
if we apply a field redefinition to the BRS operator,
any nontrivial even finite universal solution can be 
reduced to a solution such that all zeros of 
$g(w)$ lie on the unit circle.
}
In other words, we have proved that 
the coset space $\mathcal{K}$ defined by \eq (\ref{eq:coset})
is the set of all universal functions whose zeros are all on the
unit circle.

We have seen that 
the unit circle and the zeros of the universal function 
play an important role in the classification of 
the even finite universal solutions.
We consider the geometrical nature of 
these objects in the next section.

\section{Feynman propagator in Siegel gauge}

\subsection{Quadratic differentials}

In the following discussion, we take $g(w)$ to be
an even universal function defined by 
\eqs (\ref{eq:gprod}), (\ref{eq:lambdadef}) and (\ref{eq:perm}).
The kinetic operator of SFT in the Siegel gauge is 
obtained by taking the anticommutator of the BRS charge with $b_0$. 
The result is\cite{rf:TZ}
\begin{eqnarray*}
L_{v} & = & \{Q_g ,b_0 \} \\
 & = & \sum_{n} v_{n} L'_{-n} + a, 
\end{eqnarray*} 
where $v_n$ is defined by
\begin{equation}
 w g (w) = \sum_{n} v_n w^{-n+1},
\end{equation}
$L'_n = L_n + nq_n +\delta_{n,0}$ is the twisted Virasoro generator
with central charge $c=24$,and  $a$ is a constant
\footnote{Though this constant is irrelevant to
 our discussion, it would be important if we consider
the spectrum of $L_v$.}that comes from the pure
ghost term of $Q_g$.
Thus the kinetic operator is specified by the vector field
\begin{equation}
 v (w) = w g (w). \label{eq:vdefine}
\end{equation} 
This vector field satisfies the conditions,
\begin{eqnarray}
 v (w) & = & -w^2 v\left(-\frac{1}{w}\right),\label{eq:vprop1} \\
v(\pm i)              & = & \pm i,\label{eq:vprop2}
\end{eqnarray}
where we have used \eqs (\ref{eq:gcond1}) and (\ref{eq:gcond2}).
The Feynman propagator\footnote{
We omit the $b_0$ factor in the propagator.} is the inverse of 
the kinetic operator. Introducing Schwinger parameter, we have
\begin{equation}
 \frac{1}{L_{v}}  =  \int_{0}^{\infty} dt \, e^{-t L_v }.\label{eq:schwinger}
\end{equation}
We can show the Hermicity of $L_v$ using an argument 
similar to that given in the Appendix 
\ref{sec:hermite}. Thus, this propagator represents
unitary\footnote{The Wick rotation  must be taken
into account.} time evolution on the worldsheet.
The integrand of \eq (\ref{eq:schwinger}) 
acts on a primary field of dimension $d$ in the {\it twisted} CFT as 
\begin{equation}
 e^{t L_{v}} \phi (w )  e^{-t L_{v}} =
\left(\frac{d z_t (w)}{dw}\right)^{d} \phi (z_t (w)),
\end{equation} 
where $z_{t} (w)$ is a one-parameter family of 
conformal maps. It is well known that $z_t (w)$ is given
by the formula\cite{rf:LPP1,rf:LPP2}
\begin{equation}
  z_{t} (w) = e^{t v(w) \partial_w} w . \label{eq:LPPmap1}
\end{equation}
In principle, we can obtain an expression of 
 $z_t (w)$ as a formal power 
series in $t$. In order to obtain a closed expression for 
this conformal map, 
it is useful to consider the differential equation that follows
from \eq (\ref{eq:LPPmap1}) \cite{rf:LPP1,rf:LPP2}, 
\begin{equation}
 v (w ) \partial_w z_t (w) = v(z_t (w)). \label{eq:julia}
\end{equation}
When $v(w)$ is given, one can integrate the above equation and 
obtain a finite conformal map. 

In order to investigate the worldsheet geometry, 
it is useful to rewrite \eq (\ref{eq:julia}) as
an equation for
a meromorphic one form and to square this. Doing so, we obtain
\begin{equation}
 \frac{dw^2}{v(w)^2} = \frac{d z_{t}^2}{v(z_t)^2}. \label{eq:quadt} 
\end{equation}
Here, ${dz}^2 = dz dz$ is a tensor product of one forms.\footnote{Notice
that $dz^2$ does not means $dz d\bar{z}$.}
The above equation suggests the existence of a
meromorphic quadratic differential\cite{rf:quad} associated with
the open string propagator,
\begin{equation}
 \varphi(z) \, dz^2,  \quad \varphi(z) =\frac{1}{{v(z)}^2}, \label{eq:qd-def}
\end{equation}
where $z$ is the coordinate of the complex plane.

Once the quadratic differential is introduced,  
we can interpret \eq (\ref{eq:quadt}) as 
follows. Let us consider 
a region $\mathcal{R}$ in the complex plane 
such that $z_t (w)$ is single valued. Then
\eq (\ref{eq:quadt}) defines a one-parameter family of
line elements for which values of the quadratic differential are equal.
Recall that a trajectory 
of a quadratic differential is defined as an 
integral curve of line elements on $\mathcal{R}$ 
which leaves $\arg \varphi (z) dz^2 $ invariant.\cite{rf:quad} 
Therefore, the integral curve defined by \eq (\ref{eq:quadt})
is a trajectory of the quadratic differential (\ref{eq:qd-def}). 
In Appendix \ref{sec:holizontal} we 
show that $z_t (w)$ actually defines horizontal trajectories.

Among various trajectories of the
quadratic differential, the unit circle plays a special role. 
In fact, it turns out that it is always a vertical trajectory.
This can be seen by introducing the
parameterization $w=e^{i\theta}$.  Then, the value of the quadratic
differential on the unit circle is evaluated as
\begin{eqnarray}
 \frac{dw^2}{v(w)^2}  = - \frac{d\theta^2}{g(e^{i \theta})^2} \leq 0, 
\end{eqnarray}  
where we have used \eq (\ref{eq:vdefine}) and the fact that
$g(w)$ is purely real on the unit circle. 
In addition, using \eqs (\ref{eq:vprop1}) and (\ref{eq:vprop2}), we can
see that
the quadratic differential is invariant with respect to 
the BPZ inversion $z' = -1/z$:
\begin{equation}
 \frac{dz^2}{v(z)^2} = \frac{{dz'}^2}{v(z')^2}.
\end{equation}
\begin{wrapfigure}{r}{6.6cm}
\includegraphics[scale=0.7]{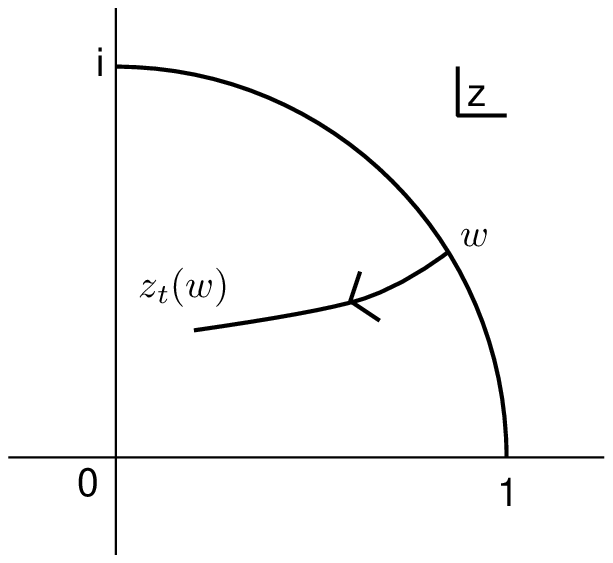}
\caption{A horizontal trajectory 
starting from unit circle.}  
\label{fig:unitevo}
\end{wrapfigure}
From the above result, it is sufficient to consider 
the region inside unit disk,
because trajectories outside the unit disk are the BPZ inversions 
of trajectories inside the region. 
Taking the starting point $w$ to be on the 
unit circle, one can uniquely determine
a horizontal trajectory that goes inside unit circle,  as illustrated
in Fig. \ref{fig:unitevo}. 
Thus we can interpret $z_t (w)$ 
as the time evolution along the horizontal
trajectory starting from the unit circle, with coordinate $w$.

The geometry of a worldsheet is determined by  
the trajectory structure of its quadratic differential. 
In particular, vertical trajectories correspond to equal time lines
of conformal fields. Because a quadratic differential is 
a coordinate independent object, its trajectory
structure is determined by the types of poles or 
zeros and the coefficients of the quadratic differential near
second order poles, which are also coordinate independent.\cite{rf:quad} 

\subsection{Regular solutions}

From \eqs (\ref{eq:gprod}) and (\ref{eq:vdefine}), 
the general form of the quadratic differential 
obtained from an even finite universal solution is given by
\begin{equation}
\varphi (z) dz^2 =
\frac{1}{\lambda^2} 
\frac{z^{4N-2}}{\prod_{k=1}^{N} (z^2-X_k)^2 (z^2 - {X_k}^{-1})^2}
 \, dz^2. \label{eq:generalquad}
\end{equation}
Now we investigate the trajectory structure of regular solutions
for which all of the $X_k$ are inside 
the unit disk.  There are $2N$ second order poles inside
the unit disk, $2N$ second order poles outside the unit disk, 
and a zero of order $(4N-2)$ at the origin. Here we list some 
features of the trajectory diagrams obtained from \eq (\ref{eq:generalquad}).

\begin{figure}[htbp]
 \begin{center}
 \includegraphics[scale=1]{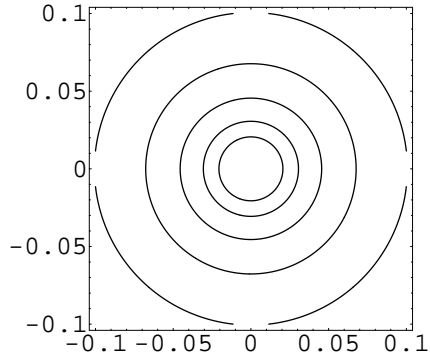}
\end{center}
\caption{Vertical trajectory around a second order pole.}
\label{fig:second}
\end{figure}

\begin{wrapfigure}{r}{6.6cm}
 \includegraphics[scale=1]{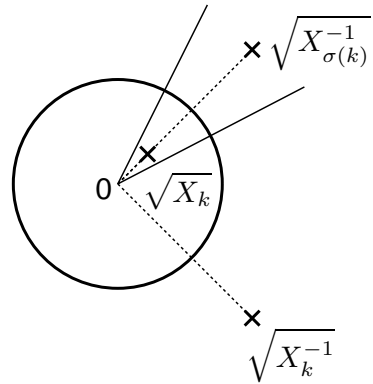}
\caption{A  pole $X_k$ is always paired with  
its inverse $X_{k}^{-1}$. From
the Hermicity condition, 
there always exists $X_{\sigma (k)}^{-1}$, which is 
complex conjugate of $X_{k}^{-1}$. }
\label{fig:pair}
\end{wrapfigure}

\begin{itemize}
 \item 
       The trajectory structure near a second order pole is
       determined by the asymptotic form of the quadratic differential
       near the pole. Near a second order pole $\sqrt{X}$,
       this asymptotic form is given by
       \begin{equation}
	\phi (z) dz^2 \sim \frac{a_{-2}}{(z-\sqrt{X})^2} dz^2,
       \end{equation}
       where $a_{-2}$ is a constant.
       If $a_{-2}$ is 
       real and positive, 
       \footnote{
       If $a_{-2}$ is  complex, the vertical trajectory 
       becomes a spiral, and the entire $z$ plane has complicated branch cuts.
       Our solutions also allow such a case. }
       the vertical trajectory is a closed curve surrounding
       the pole (see Fig.\ \ref{fig:second}).  
       For convenience, we consider
       the cases in which all poles have positive coefficient.

 \item 
       For a second order pole $\sqrt{X_k}$ inside the unit disk,
       there always exists the other pole $\sqrt{X_{\sigma (k)}^{-1}}$ outside 
       the unit disk (Fig.\ \ref{fig:pair}). 
       This follows from the Hermicity condition
       (\ref{eq:perm}). Such pairs of poles play an important role in
       determining the worldsheet geometry.
\end{itemize}

\begin{figure}[htbp]
 \parbox{\halftext}{
\centerline{\includegraphics[width=5.8cm]{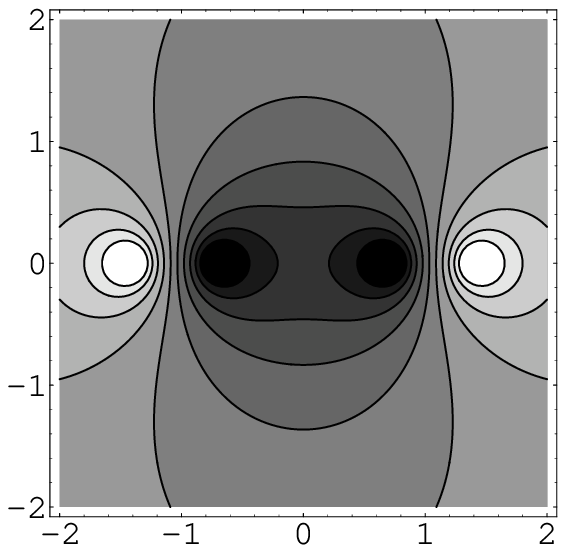}}
\caption{Vertical trajectories of the $N=1$ function with
$X_1 = 1/2$. From the Hermicity condition, the poles 
must be real or purely imaginary.
The shade of a region represents the worldsheet `time'
$\tau=\mathrm{Re}(\rho)$, with increasing brightness
corresponding to increasing time.
}
\label{fig:gn1}}%
\hfill
 \parbox{\halftext}{
\vspace{-40pt}
\centerline{\includegraphics[width=5.8cm]{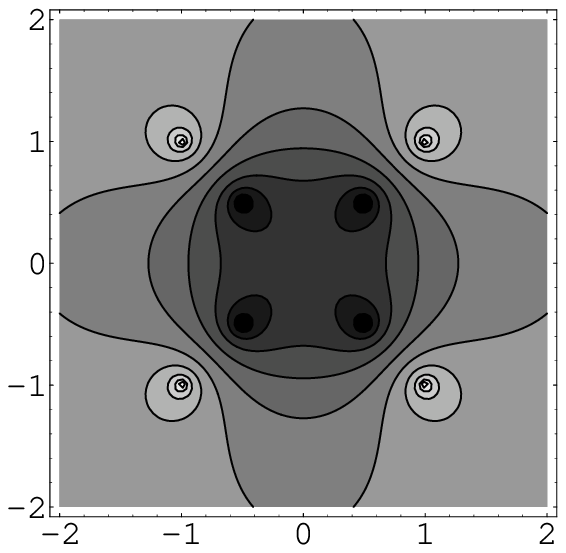}}
\caption{Vertical trajectories of the $N=2$ function with
$X_1 = i/2$. The four poles inside the unit disk 
correspond to
the two square roots of $X_1$ and the 
two square roots of $\overline{X_1}$.}
\label{fig:gn2}}
\end{figure}

In order to draw  vertical trajectories in the $z$ plane, 
it is convenient to introduce the flat (or `distinguished') 
coordinate of the quadratic differential (\ref{eq:generalquad}):
\begin{equation}
 d \rho ^2 = \varphi (z) dz^2.\label{eq:rhocoord}
\end{equation}
Integrating the above equation, we find
\begin{equation}
 \rho = \int^{z} \sqrt{\phi (z')} \, dz' = \Phi (z) .\label{eq:distinguished}
\end{equation}
where $\rho$ is defined up to its sign and an additive constant.\footnote{Using $\Phi (z)$ and \eq (\ref{eq:quadt}), we 
can obtain the formula $z_{t} (w) =  \Phi^{-1} (\Phi(w) +t)$,
which is equivalent to 
the formula in Ref.~\citen{rf:projector}. }
Because a vertical trajectory in this flat coordinate
is a vertical straight line (i.e. a curve that satisfy
$\mathrm{Re} \, \rho = $ const.\ )
a vertical trajectory in the $z$ plane is given by the condition
\begin{equation}
 \mathrm{Re} (\Phi (z)) = r.  \quad  (r \in \mathbb{R})
\end{equation}
\begin{wrapfigure}{r}{6.6cm}
\begin{center}
 \includegraphics[width=5.8cm]{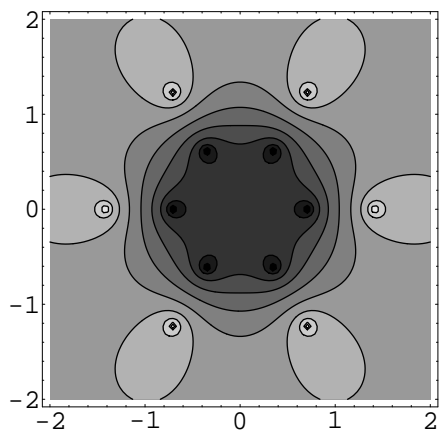} 
\caption{Vertical trajectories of $N=3$ function with
$X_1 =  e^{2\pi i /3 } /2 ,X_2 =1/2$. 
In this case, $X_2$ forced to be real. }
\label{fig:gn3}
\end{center}
\end{wrapfigure}
Here we give some examples.
Using the general expression 
of the universal function (\ref{eq:generalquad}),
we plot vertical trajectories 
of $N=1$ (Fig.\ \ref{fig:gn1}),
$N=2$ (Fig.\ \ref{fig:gn2}) and $N=3$ (Fig.\ \ref{fig:gn3}) functions. 
The positions of the poles are chosen
so as to satisfy \eq (\ref{eq:perm}). 

\subsection{%
Example of an open string worldsheet with $N=3$}
\label{subsec:N3}

As seen in the figures  \ref{fig:gn3} and \ref{fig:gn2}, 
an $N\geq2$ diagram has many second order poles in general.
It looks like a worldsheet of interacting strings, because
a second order pole corresponds to an incoming or outgoing source
of an open string state in the ordinary open string CFT.
However, these graphs must describe the propagation of
a single string, as we consider only single propagator. 
As we see below, it turns out that
most poles are {\it dummies}, and only two pairs of poles 
couple with string states.  

In order to specify the region corresponding to
an open string in the trajectory diagram, 
we start from an alternative derivation of the quadratic
differential based on a gauge invariant treatment. Suppose that 
a BRS charge with a universal function $g(w)$ is obtained from
ordinary BRS charge $Q_B$ by twisted conformal transformation
\begin{equation}
  \mathcal{N} Q_g = U'_{f} Q_B {U'_{f} }^{-1},\label{eq:Qconformal}
\end{equation}
where $U'_f$ is a finite conformal transformation with the map $z=f(w)$ 
in the twisted CFT, and  $\mathcal{N}$ is a real constant.  
Then, using the fact that the BRS current transforms like a weight two operator
, up to the pure ghost term\cite{rf:TZ},
and using the relation $J'_B (w) = w J_B (w)$, where $J'_B (w)$
is the twisted BRS current,   
we can compute the $J_B$ part of the right-hand side of \eq (\ref{eq:Qconformal})  as
\begin{eqnarray}
 \oint_{\gamma} dw \, U'_f J_B (w) {U'_f}^{-1} & = &
  \oint_{\gamma} dw \, \frac{w}{z} 
\left( \frac{dz}{dw}\right)^2 J'_B (z) \nonumber  \\
& = &  \oint_{\gamma} dz \, \frac{w}{z} 
\left( \frac{dz}{dw}\right) J'_B (z). \label{eq:BRStransf}
\end{eqnarray}
In the second line of the above equation, we assume that $z$ winds once 
on the unit circle. 
Comparing this expression to the left-hand side
of \eq (\ref{eq:Qconformal}), we obtain
\begin{equation}
 \mathcal{N}  g (z) =\frac{w}{z} \frac{dz}{dw}, \label{eq:diffeq}
\end{equation}
or, rewriting the above equation as a one form and squaring, we obtain
\begin{equation}
 \frac{dz^2}{v(z)^2} =  \mathcal{N}^2 \frac{dw^2}{w^2}.\label{eq:qdfromQ}
\end{equation}
The right-hand side of \eq (\ref{eq:qdfromQ}) is 
the quadratic differential in the disk coordinate $w$.
Indeed, we can obtain \eq (\ref{eq:rhocoord}) from \eq (\ref{eq:qdfromQ})
by using the  coordinate transformation $w = e^{\mathcal{N} \rho}$.
Therefore, the 
conformal transformation $z=f(w)$ maps a 
disk worldsheet into 
a certain region in the $z$ plane, and \eq (\ref{eq:qdfromQ}) represents
a coordinate transformation of the quadratic differential. 

\begin{figure}[htbp]
\begin{center}
 \includegraphics[scale=0.8]{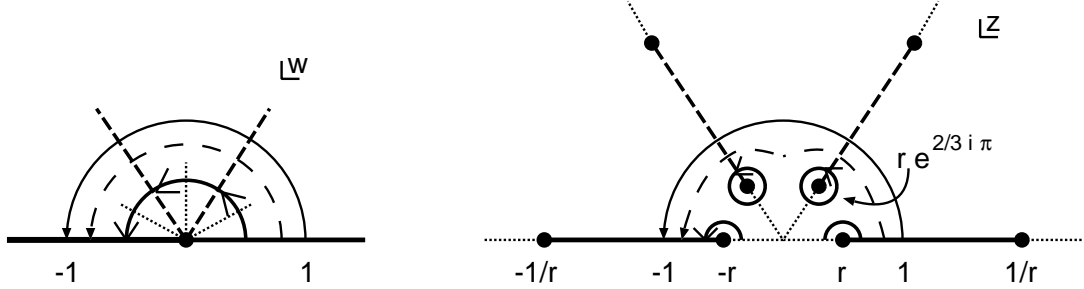} 
\caption{Conformal map of the $N=3$ solution. Four poles inside upper
 half $z$ planes are `dummies'.}
\label{fig:l2sol}
\end{center}
\end{figure}

As an example, we derive the 
conformal map associated with the one parameter family of  $N=3$
solutions (which includes the case of  Fig.\  \ref{fig:gn3})
\footnote{This solution is  $l=3$ case of 
the Kishimoto--Takahashi solution found in Ref.\ \cite{rf:KT}. },
i.e. 
\begin{equation}
X_1 = r^2  e^{\frac{2}{3} i \pi } ,\ X_2 =r^2,\  X_3 = r^2  e^{-\frac{2}{3} i \pi }
\quad (0 \leq r < 1).
\end{equation}
From \eq (\ref{eq:generalquad}), the quadratic differential is given
by
\begin{equation}
 \phi (z) dz^2 = \frac{(1+r^6)^4}{r^{12}}
\frac{z^{10} dz^2}{(z^6 -r^6)^2(z^6- r^{-6})^2},
\end{equation}
and the differential equation (\ref{eq:diffeq}) becomes 
\begin{equation}
 -\frac{(1+r^6)^2}{r^6} \frac{z^5 dz}{(z^6-r^6)(z^6-r^{-6})}
= \mathcal{N} \frac{dw}{w}.\label{eq:defN2}
\end{equation}
We can obtain an explicit expression of $z=f(w)$ by 
integrating \eq (\ref{eq:defN2}). Note that there are
two unknown constants, $\mathcal{N}$ and an integration 
constant. They are determined from the condition that
{\it the entire unit circle be fixed with respect to
the conformal map.} 
This condition ensures the gluing rule of states in the SFT with
BRS charge $Q_g$, because SFT vertices are defined by the
overlap conditions on the unit circle.
In this example, these constants are fixed as follows: 
\begin{itemize}
 \item $\mathcal{N}$ is fixed so that the
closed path around the unit circle winds once in both
$w$ and $z$ planes. 
       This condition is already assumed in \eq (\ref{eq:BRStransf}).
       Then, from \eq (\ref{eq:defN2}) we obtain
       $\mathcal{N} = (1+r^6)/(1-r^6)$. 
 \item The integration constant in \eq (\ref{eq:defN2}) is fixed 
       by the condition $w=\pm i \Rightarrow z=\pm i$, which fixes
       the open string midpoint. This condition yields the operator $U'_f$ 
       generated by the midpoint preserving operator 
       $K'_{2m} =L'_{2m} -L'_{-2m}$ \cite{rf:CSFT}.  
\end{itemize}
Fixing these constant, 
we find  the local expression for the conformal map 
\begin{equation}
 z = f(w)= \left(\frac{w^6+ r^6}{r^6 w^6+1}\right)^\frac{1}{6}. \label{eq:n2finitef}
\end{equation}
Because we already know that the equal time
lines in the $z$ plane are always on the vertical trajectory depicted in
Fig.~\ref{fig:gn3}, all we need is to identify the boundaries and 
the image of the upper half $w$ plane. This can be 
accomplished easily  by using
(\ref{eq:n2finitef}).   Figure \ref{fig:l2sol} shows how the upper half of $w$
plane is mapped into the $z$ plane. 
In both $z$ and $w$ coordinates, open string
worldsheets constitute upper half planes, while
open string boundaries in the $z$ plane
consist of two line segments connecting second order poles on the real axis.
The small semicircle around $w=0$ goes around 
two second order poles on the real axis and two
spurious second order poles in the upper half $z$ plane.
Contrastingly, a semicircle near $|w|=1$ is mapped 
to a single curve near $|z|=1$, because it does not meet brunch cuts in the 
$w$ plane.     
Similar results are obtained for different $N$ cases: 
most second order poles are dummies, and an open string boundary
is a vertical trajectory that connects two second order poles
on the real axis.

Though we have given an example of
the identification of an open string worldsheet explicitly,
such analysis is not necessary in determining the singularity of a
solution, as
it depends only on the local structure of the quadratic differential. 

\subsection{Singular solutions}
\label{subsec:singular}

\begin{figure}[htbp]
\centerline{\includegraphics[scale=1.2,clip]{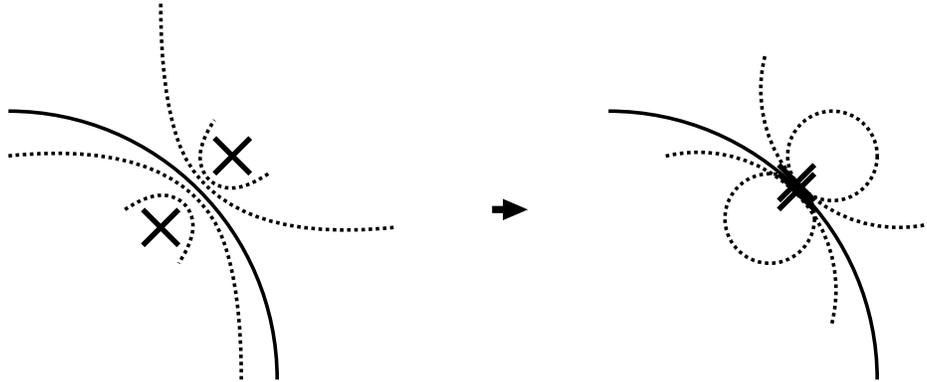}}
\caption{A fourth order pole on the unit disk
can be regarded as coincident second order poles. 
}
\label{fig:pinch}
\end{figure}

\begin{figure}[htbp]
 \parbox{\halftext}{
\centerline{\includegraphics{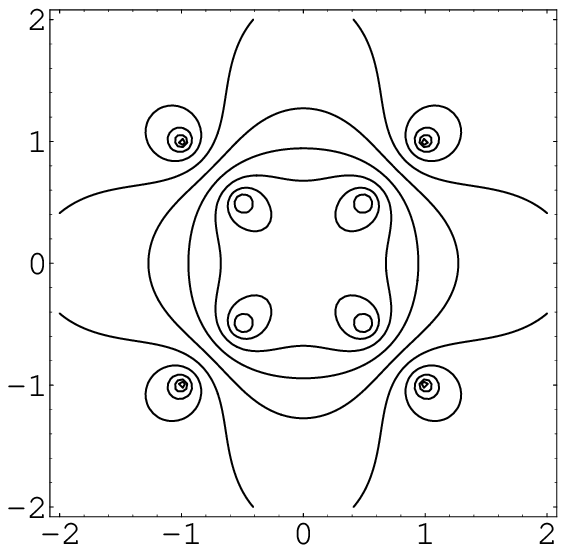}}
\caption{Vertical trajectories of the $N=2$ solution with
$X_1 = e^{i\pi/2} / 2 $. Two second order poles are 
situated across from each other 
with respect to the unit circle.}
\label{fig:rhalf}}
\hfill
 \parbox{\halftext}{
\centerline{\includegraphics{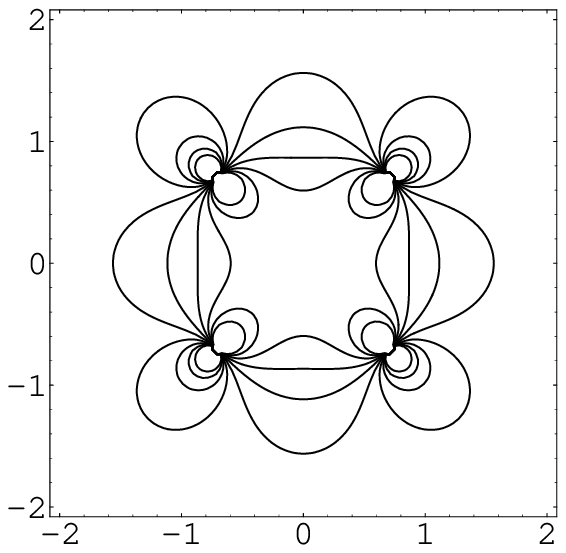}}
\caption{Vertical trajectories of the $N=2$ solution with
$X_1= e^{i\pi/2}$. A pair of second order poles
becomes a single fourth order pole.}
\label{fig:r1}}
\end{figure}

In the case of singular solutions,
any universal function can be taken
to have all of its  zeros on the unit circle.
In the language of quadratic differentials, these zeros
correspond to second order poles on the unit circle, as seen from \eq
(\ref{eq:generalquad}).   
Because a pole $\sqrt{X_k}$ inside the unit disk 
is always paired with another pole $\sqrt{ \overline{X_{k}^{-1}} }$ outside 
the unit disk, 
they coincide on the unit circle and become a single
fourth order pole.  
Thus we conclude
that {\it the quadratic differential associated with a singular 
even finite universal solution
must have  poles of fourth or more higher order
on the unit circle.
}
Figure  \ref{fig:pinch} illustrates the local trajectory structure 
near the unit circle.  Figures  \ref{fig:rhalf} and \ref{fig:r1}
are trajectory diagrams of the $N=2$ solutions. 
The result obtained above implies that {\it all} 
singular even finite solutions correspond
to the no open string vacuum, because the local structure of
the quadratic differential around
a pole of fourth or more higher order
is quite different from that of a pole of second order, which
can be mapped to a flat strip. This implies that the Feynman propagator  
around a singular solution never couples with open string states. 
To make this observation more
evident, we must analyze the spectrum or the cohomology of the BRS charge
for singular solutions.

\section{Conclusions}

In this paper we have investigated a class of 
universal solutions specified by 
even polynomial universal 
functions and showed
that a solution becomes nontrivial when the zeros of
the universal function are on the unit circle. 
Furthermore, we have found
that the universal function associated with each solution 
gives a meromorphic quadratic differential
on the complex plane, and 
that the solution becomes nontrivial
when second order poles coincide on the unit circle. 
Such a relation between nontrivial universal solution and
poles on the unit circle has been pointed out by 
Drukker\cite{rf:Drkr1,rf:Drkr2}. 
Our result confirms his conjecture
in the case of  even order polynomial universal
functions. 
Now it becomes clear that the universal solution uniquely
determines the string propagation around itself, and the geometrical
nature of the worldsheet swept by the string
are entirely encoded in the universal
function. Furthermore, we know that a universal solution
can be formally expressed as a pure gauge solution whose 
gauge group element is uniquely specified by the universal 
function. Therefore, these facts 
imply that the gauge group of SFT contains 
rich degrees of freedom that correspond to 
various string worldsheets. 

We conjecture that all nontrivial solutions considered in this
paper yield tachyon condensation and the disappearance of open strings,
because the propagators never couple to open string states. 
Though we have obtained a wide variety of nontrivial solutions,
according to Sen's conjecture,
the stable vacuum of the tachyonic potential
must be unique. If Sen's conjecture holds, 
these solutions must be related to each other by some transformations.
Therefore it is important to investigate 
whether the nontrivial solutions are equivalent.
Furthermore, to confirm the conjecture that nontrivial universal 
solutions yield closed string propagation, we must calculate 
closed string amplitudes using these solutions. 
It has been suggest that a zero momentum dilaton 
lives on the shrunken boundary\cite{rf:vsftghost,rf:Drkr1}.
It would be interesting to calculate this amplitude 
explicitly in same manner as Ref.~\citen{rf:TZ2}.

Although we have investigated even
finite universal solutions only, 
it would be interesting to consider other cases.
First, we must consider 
nonpolynomial universal functions, because they appear 
as the inverse elements of 
polynomial universal functions with respect to
the Abelian multiplication discussed in \S \ref{sec:universal}.  
Another example of
a nonpolynomial type solution  
that yields a pure ghost BRS charge 
is given in Ref.~\citen{rf:Drkr2}. In addition, polynomial-type
solutions with odd parts also must be considered.

\section*{Acknowledgements}
I would like to thank T.~Takahashi for many 
helpful discussions. I would also like to thank 
Y.~Igarashi, K.~Itoh, F.~Katsumata  and D.~Belov
for valuable discussions,
and H.~Itoyama, M.~Sakaguchi, S.~Tanimura
and Y.~Yasui for helpful comments.  
This work is supported in part
by the 21 COE program ``Constitution of wide-angle 
mathematical basis focused on knots''.

\appendix

\section{Hermicity of the BRS Charge}
\label{sec:hermite}

Let $g(w)$ be a universal function in $\mathcal{F}$ with a finite Laurent
expansion. It can be expanded as
\begin{equation}
 g(w) = \sum_{n=-M}^{M} g_n w^{-n},  \label{eq:hermg}
\end{equation}
where $M$ is a positive integer.
First, we investigate the Hermicity condition of $Q(g)$.
A mode expansion of $J_{B} (w)$ is given by
\begin{equation}
 J_{B} (w) = \sum_{n} Q_n w^{-n-1}.\label{eq:JBexpand}
\end{equation}
Then, using \eqs (\ref{eq:hermg}), (\ref{eq:JBexpand}) and (\ref{eq:Q(F)def}), 
the mode expansion of $Q(g)$ is found to be 
\begin{equation}
 Q(g) = \int_{0}^{2 \pi} \frac{d\theta} {2 \pi} 
\sum_{n=-M}^{M}
\sum_{m=-M}^{M} g_{n} Q_{m} e^{-i(n+m)\theta },\label{eq:Q(g)mode} 
\end{equation}  
where we have rewritten the contour integral along $\gamma$ into 
the form of a real
integral over $\theta$ by using $w=e^{i \theta}$.  
Next, using ${Q_n}^{\dagger} = Q_{-n}$, 
it is easy to see that the right-hand side of
 \eq (\ref{eq:Q(g)mode}) is Hermite if
\begin{equation}
 \overline{g_n} = g_{-n}.\label{eq:realgn}
\end{equation}
The above condition is equivalent to 
the situation that $\overline{g(w)} = g(w)$
on the unit disk. Indeed, using \eqs (\ref{eq:hermg}) and
(\ref{eq:realgn}), we can show
\begin{eqnarray}
 \overline{g(w)} & = & \sum_{n=-M}^{M} \overline{g_n}\, 
 (\overline{w})^{-n}  \nonumber \\ 
 & = & \sum_{n=-M}^{M} g_{-n} w^{n} \nonumber \\ 
 & = & g(w),
\end{eqnarray}
where we have used $\overline{w}= 1/w$ in the second line. In similar
way,  we can show that $C(f)$ is Hermite if $\overline{f(w)} = w^4 
f(w)$ is satisfied  on the unit disk. When
\begin{equation}
 f(w) = \frac{(\partial g (w) )^2}{g(w)},
\end{equation} 
we can show that $\overline{f(w)} = w^4 
f(w)$ is  satisfied using \eq (\ref{eq:realgn}). Thus we have shown 
that $Q_g =Q(g) -C(f)$ is Hermite if $g(w)$ is real on the 
unit disk.

\section{The Ghost Current Operator}
\label{seq:qh}

Here we evaluate the Laurent expansion of $h(w)= \log g(w)$ 
on the unit circle, 
because all fields and functions in this paper are defined
around the unit circle.
First, recall that even finite universal functions satisfy
$g(w)=\overline{g(w)}$ on the unit circle. Moreover, using
the cyclic decomposition (\ref{eq:cyclicdecomp}), we can write
\eq (\ref{eq:decomposeg}) as
\begin{equation}
 g(w) =
\begin{cases} \displaystyle
\prod_{k=1}^{N-1}
    \left| g_{X_k} (w) \right|^2   g_{X_N } (w) & (N \in 2\mathbb{N} -1),
\label{eq:greals} \\
\displaystyle \prod_{k=1}^{N  } 
    \left| g_{X_k} (w) \right|^2 & (N \in 2\mathbb{N} ),   
\end{cases}
\end{equation}
on the unit circle, where $g_{X_k} (w)$ is 
defined by \eq (\ref{eq:gXk}), and  $X_N$  is real in the odd $N$ case. 
Then, using analysis similar to that given
in Appendix of Ref.~\citen{rf:TT}, 
$g_{X_k} (w)$ is  positive on the unit circle 
when $X_k$ is real.
Therefore,  $g(w)$ is positive on the unit disk in 
both cases of \eq (\ref{eq:greals}), and 
$\log g(w)$ takes real values.
Then, applying a procedure similar to that given
in Ref.~\citen{rf:TT}, we obtain
\begin{equation}
 h (w) = - \sum_{k=1}^{N}  \log\left(1+X_k \right)^2 -
\sum_{n=1}^{\infty} \frac{
\left(
\sum_{k=1}^{N} X_{k}^{n} \right) }{n} \left(
w^{2n} + {w^{-2n}} \right).\label{eq:hexpand}
\end{equation}
Indeed, 
we can see that \eq (\ref{eq:hexpand}) is  real 
on the unit circle using  \eq (\ref{eq:perm}).

The ghost current operator is
similarly expanded as
\begin{equation}
  q(h) = - \sum_{k=1}^{N}  \log\left(1+X_k \right)^2 \ (q_0 +1) -
\sum_{n=1}^{\infty} \frac{
\left(
\sum_{k=1}^{N} X_{k}^{n} \right) }{n} \left(
q_{2n} + q_{-2n} \right),\label{eq:qhexpand}
\end{equation}
where $J_{gh} (w) = \sum_{n} q_n w^{-n-1}$. Using the above expansion
and $\left[q_m, q_n \right] = m \delta_{m+n,0}$, we obtain
\begin{eqnarray}
 \left[q^{(+)} (h),q^{(-)} (h)\right] & = &
2 \sum_{n=1}^{\infty} \frac{
\left(\sum_{k=1}^{N} X_{k}^{n}    \right)^2}{n} \nonumber\label{eq:qpmcom} \\ 
 & = & -2 \sum_{ k=1}^{N}  \sum_{l=1}^{N}
 \log X_k X_{l}.
\end{eqnarray} 
Here, in the second line of \eq (\ref{eq:qpmcom}), 
we have used the fact that $X_k$ is inside the 
unit disk. 

Using \eq (\ref{eq:qhexpand}),
it is easy to demonstrate that the Hermicity of BRS charge. 
It is clear that
${q(h)}^{\dagger} = - q (h)$ is satisfied, because
${q_{2n}}^{\dagger}= -q_{-2n}$ and coefficients 
in \eq (\ref{eq:qhexpand}) are real.
Therefore we write the BRS charge as
\begin{equation}
 Q_g = e^{q(h)} Q_B e^{-q(h)},
\end{equation}
then the Hermicity of $Q_g$  
follows from that of $Q_B$.

\section{Finite Confocal Map and Horizontal Trajectory}
\label{sec:holizontal}

Here we show that 
the conformal map $z_t (w)$ defined by \eq (\ref{eq:LPPmap1})
naturally defines horizontal trajectories. 
First, integrating \eq (\ref{eq:julia}), we obtain
\begin{equation}
 \Phi (z_t (w)) = \Phi (w) + t, \label{eq:2-phi}
\end{equation}
where $\Phi (z)$ is defined in \eq (\ref{eq:distinguished}).  
The constant $t$ in the right-hand side of 
\eq (\ref{eq:2-phi}) is determined by differentiating this equation
with respect to $t$ and using \eq (\ref{eq:julia}).
Let us consider a path starting from a point on the 
unit circle.  We 
 introduce the parameterization $w=e^{i \theta}$. Plugging this
into \eq (\ref{eq:2-phi}), we have
\begin{equation}
 \Phi (z_t (e^{i \theta})) = \Phi (e^{i\theta}) + t. \label{eq:2-phi-theta}
\end{equation}
Next, we show that $\Phi (e^{i\theta})$ is purely imaginary. 
From \eqs (\ref{eq:distinguished})  and (\ref{eq:qd-def}), we have
\begin{eqnarray}
 \frac{d}{d\theta} \Phi (e^{i \theta}) 
& = & \frac{i e^{t \theta }}{v(e^{i \theta})} \nonumber \\
& = & \frac{i}{g(e^{i\theta})}.
\end{eqnarray}
Then, from the above equation, we find that 
\begin{equation}
 \frac{d}{d\theta}\  \mathrm{Re} \ \Phi (e^{i \theta}) =0,
\end{equation}
because $g(e^{i \theta})$ is real, 
as shown in Appendix \ref{sec:hermite}.  Thus the real part of 
$\Phi (e^{i \theta})$
is a constant. Furthermore, we can set this constant to zero
because $\Phi (z)$ is  defined only up to an additive constant. Therefore
\eq (\ref{eq:2-phi-theta}) can be expressed as
\begin{equation}
 \Phi (z_t (e^{i \theta})) = i f(\theta) + t, 
\end{equation} 
where $f(\theta)$ is a real valued function.
Note that the right-hand side of this equation corresponds 
to the flat coordinate $\rho$ introduced in \eq (\ref{eq:distinguished}).
Let us start on the unit circle and move inside the unit circle with
$t$ decreasing and $\theta$ fixed. A curve obtained in this way
is a 
horizontal trajectory in the $\rho$ coordinate, and it is also
horizontal in the $z$ coordinate.

\end{document}